\documentclass[twocolumn]{aastex631} 

\usepackage{bm}
\usepackage{url}
\usepackage{soul}

\begin{document}

\title{High-energy Neutrinos from the Inner Circumnuclear Region of NGC~1068} 

\author[0000-0002-5387-8138]{K.~Fang}
\affiliation{Department of Physics, Wisconsin IceCube Particle Astrophysics Center, University of Wisconsin, Madison, WI, 53706 }

\author[0000-0001-5357-6538]{Enrique Lopez Rodriguez}
\affiliation{Kavli Institute for Particle Astrophysics and Cosmology, Stanford University, Stanford, CA 94305}

\author[0000-0001-6224-2417]{Francis Halzen}
\affiliation{Department of Physics, Wisconsin IceCube Particle Astrophysics Center, University of Wisconsin, Madison, WI, 53706} 

\author[0000-0001-8608-0408]{John S. Gallagher}
\affiliation{Department of Astronomy,  University of Wisconsin, Madison, WI, 53706}

\date{\today}

\begin{abstract}
High-energy neutrinos are detected by the IceCube Observatory in the direction of NGC~1068, the archetypical type II Seyfert galaxy. The neutrino flux, surprisingly, is more than an order of magnitude higher than the $\gamma$-ray upper limits at measured TeV energy, posing tight constraints on the physical conditions of a neutrino production site. We report an analysis of the sub-millimeter, mid-infrared, and ultraviolet observations of the central $50$ pc of NGC~1068 and suggest that the inner dusty torus and the region where the jet interacts with the surrounding interstellar medium (ISM) may be a potential neutrino production site. Based on radiation and magnetic field properties derived from observations, we calculate the electromagnetic cascade of the $\gamma$-rays accompanying the neutrinos. When injecting protons with a hard spectrum, our model may explain the observed neutrino flux above $\sim 10$~TeV. It predicts a unique sub-TeV $\gamma$-ray component, which could be identified by a future observation. Jet-ISM interactions are commonly observed in the proximity of jets of both supermassive and stellar-mass black holes. Our results imply that such interaction regions could be $\gamma$-ray obscured neutrino production sites, which are needed to explain the IceCube diffuse neutrino flux. 
\end{abstract}

\section{introduction}

An excess of high-energy neutrinos with a global significance of 4.2~$\sigma$ was identified in the direction of the active galaxy NGC~1068 by the IceCube Observatory \citep{IceCubeNGC1068}. The neutrino energy flux observed by IceCube is more than an order of magnitude higher than the upper limits on the $\gamma$-ray flux at TeV energies, suggesting that the neutrino emission site must be highly $\gamma$-ray-obscured. This agrees with the indication, based on a comparison of the diffuse extragalactic fluxes, that neutrinos originate in cosmic environments that are optically thick to GeV–TeV $\gamma$-rays \citep{2016PhRvL.116g1101M, 2022ApJ...933..190F}.

NGC~1068 at a distance of 14.4~Mpc (1\arcsec = 60 pc) is the brightest Seyfert galaxy \citep{1909LicOB...5...71F, 1997Ap&SS.248....9B}.  
The mass of the central black hole is estimated as $(0.8-9.5)\times 10^7\,M_\odot$ \citep{2003A&A...398..517L,2015ApJ...802...98M}. The bolometric luminosity of the active galactic nucleus (AGN) is  estimated to be $L_{\rm bol}=5.02^{+0.15}_{-0.19}\times 10^{44}\,\rm erg\,s^{-1}$ based on mid-infrared to sub-mm spectral modeling \citep{LR2018}. 
No short- or long-term line variability has been found in the X-ray data, suggesting that a good fraction of the emission originates from  regions well outside of the parsec-scale dusty and molecular torus \citep{2015ApJ...812..116B, 2021A&A...649A.162G}. 

Due to the high photon opacity to TeV $\gamma$-rays, a possible coronal region in the proximity of the central black is an appealing site for the neutrino production. Models of neutrino production in the corona of the supermassive black hole have been explored for example in \citet{2020ApJ...891L..33I, PhysRevLett.125.011101, 2021ApJ...922...45K, 2021Galax...9...36I, 2022icrc.confE.993A, 2022arXiv220702097I, 2022ApJ...939...43E}. The models generally require that the emission regions are located within $\sim30-100$ Schwarzschild radii \citep{2022ApJ...941L..17M}.

An extraordinary component of NGC~1068 is its bright and complex circumnuclear region. The AGN is hidden behind a nearly edge-on dusty and molecular disk at parsec scales extending up to 10-200~pc that is misaligned with the spiral galaxy \citep{2016ApJ...823L..12G,  2022Natur.602..403G}.  
Radio and molecular line observations of the circumnuclear region \citep{2019A&A...632A..61G, 2019ApJ...884L..28I} suggest that the radiation pressure drives molecular outflows in the inner region of the gas disk ($R\lesssim 3$~pc). In addition, NGC 1068's kiloparsec-scale, steep-spectrum radio jet interacts with the interstellar medium (ISM) in the central tens of pc. Along the jet axis, several nuclear radio sources are observed, referred to as radio emission ``knots" as shown in Figure \ref{fig:fig1} \citep{2004ApJ...613..794G, 2020ApJ...893...33L}.

Star formation is observed both inside the circumnuclear disk within $\sim 200$~pc and at a circumnuclear starburst ring at kpc scale. The star formation regions may contribute to the $\gamma$-ray emission at 1-100~GeV \citep{2014ApJ...780..137Y}. Though they are mostly optically thin to TeV $\gamma$-rays and thus cannot be the site where neutrinos are produced.

In this paper, we investigate the neutrino production in the central $50$ pc of NGC~1068 where the jet interacts with the ISM and in the dusty torus. We analyze the sub-millimeter (mm), mid-infrared (MIR), and ultraviolet (UV) observations of the radio knots to obtain the spectrum and energy density of the radiation fields. We also use sub-mm polarization measurements to infer the strength of the magnetic fields (B-fields). We find that the intense IR and optical fields of the knots may attenuate TeV $\gamma$-rays that accompany the production of high-energy neutrinos. The region, however, is not sufficiently thick to 10-100~GeV $\gamma$-rays, resulting in a unique sub-TeV $\gamma$-ray component that may be revealed by future observation. We present the observation and analysis of the radio knots in Section~\ref{sec:knotObs}. We investigate the production of high-energy neutrinos and attenuation of TeV $\gamma$-rays in Section~\ref{sec:nu}. Finally, we discuss and conclude in Section~\ref{sec:dis}.

\section{Observation of the jet-interacting knots}\label{sec:knotObs}

\begin{figure*}  
    \centering
   \includegraphics[width=0.99\textwidth]{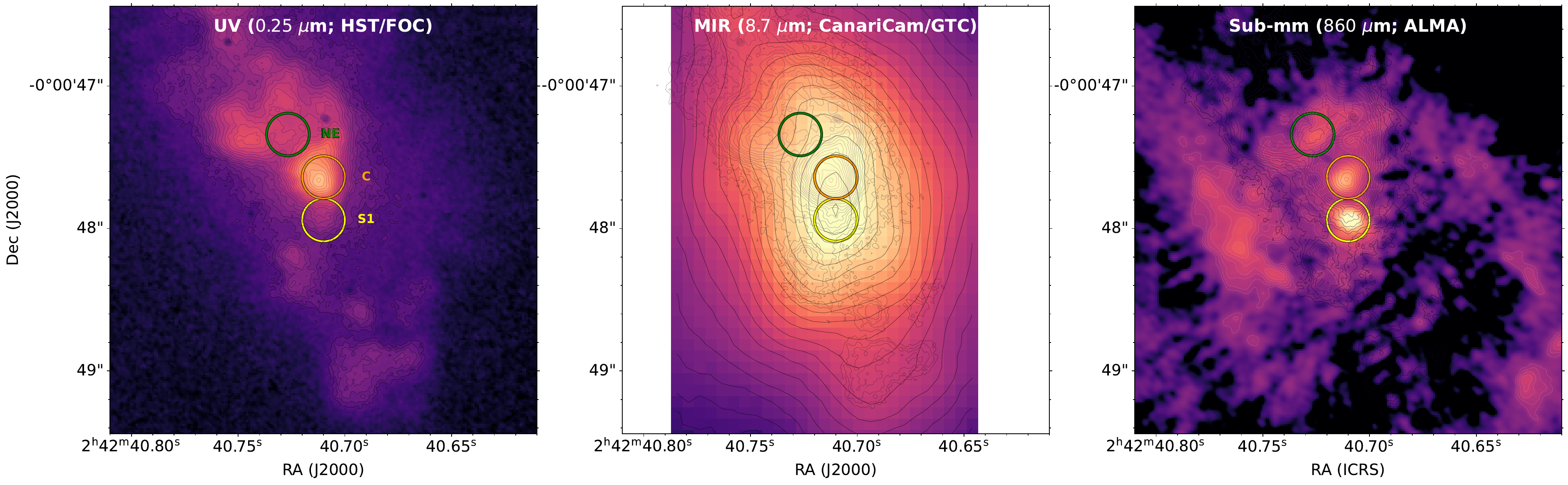}
    \caption{
    \label{fig:knots} The multi-wavelength emission of the central $180\times180$ pc$^{2}$ region of NGC~1068. The total intensity maps at $0.25~\mu$m from \textit{HST}/FOC (left; \citealt{Kishimoto1999, Barnouin}), $8.7~\mu$m from CanariCam/GTC (middle; \citealt{LR2016}), and $860~\mu$m from ALMA (right;  \citealp{2020ApJ...893...33L}) are shown. The contours start at $3\sigma$ and increase as $1.5^{n}\sigma$, where $n=1, 1.5, 2, 2.5, \dots$ , and $\sigma$ is the noise of the background regions of each observation. The circles show the aperture size of $0\farcs3$ ($18$ pc) used to compute the photometry of the `S1' (yellow), `C' (orange), and `NE' (green) knots.} \label{fig:fig1}
\end{figure*}

\begin{deluxetable*}{lccccccc}
\centering
\tablecaption{Photometry of the knots.
\label{tab:table1} 
}
\tablecolumns{6}
\tablewidth{0pt}
\tablehead{\colhead{Knot} & 	\multicolumn{6}{c}{Fluxes [Jy]}  \\ 
    &   \colhead{$0.25~\mu$m}  & \colhead{$8.7~\mu$m} & \colhead{$10.3~\mu$m} & 
\colhead{$11.3~\mu$m} &	\colhead{$11.6~\mu$m} &  \colhead{$860~\mu$m} } 
\startdata
S1 (core)      &   $(2.76\pm0.11)\times10^{-4}$    &   $7.9\pm0.8$ &   $7.2\pm0.8$ &   $10.4\pm1.2$    &   $11.2\pm1.2$	& $(1.69\pm0.17)\times10^{-2}$ \\
C      &   $(2.89\pm0.01)\times10^{-3}$    &   $4.5\pm0.4$ &   $6.1\pm0.6$ &   $8.4\pm0.7$    &   $7.2\pm0.7$	& $(6.47\pm0.65)\times10^{-3}$  \\
NE      &   $(6.37\pm0.03)\times10^{-4}$    &   - &   $0.6\pm0.1$ &   -    &   -	& $(3.23\pm0.10)\times10^{-3}$  \\
\enddata
\end{deluxetable*}

Several bright knots are found within the central $\sim50$ pc of NGC~1068 observed at radio observations \citep{2004ApJ...613..794G}. Specifically, knot `S1' is identified as the core of NGC~1068. Knot `C' is identified as the location of the interaction between the jet and a giant molecular cloud (GMC) at $\sim30$ pc north from `S1'. Knot `NE' is identified as the interaction of the jet with the ISM at $\sim50$ pc north-east from `S1' after the jet bent to an angle of $45^{\circ}$ east of north due to the interaction in knot `C'. Figure \ref{fig:fig1} shows the knots `S1', `C', and `NE' over the total intensity observations used in this work. 

We compute the energy spectral distribution (SED) of these knots at the highest angular resolution observations using imaging polarimetric observations at sub-mm, MIR, and UV wavelengths. The highest angular resolution observations are needed to ensure that the knots are spatially resolved, so the SED of each knot can be studied  without the contamination of the extended and diffuse emission of the circumnuclear regions of NGC~1068. The polarimetric observations are needed to estimate the B-field strength of these regions. For the sub-mm observations, we use the $860~\mu$m ($348.65$ GHz) imaging polarimetric observations at an angular resolution of $0\farcs07$ ($4.2$ pc) by ALMA \citep{2020ApJ...893...33L}. For the MIR observations, we use the $8.7-11.6~\mu$m ($3.4-2.6\times10^{4}$ GHz) imaging- and spectro-polarimetric observations at an angular resolution of $0\farcs3$ ($18$ pc) with CanariCam on the 10.4-m Gran Telescopio CANARIAS (GTC) \citep{LR2016}. For the UV observations, we use the F253M UV filter with a center wavelength of $0.25~\mu$m ($1.2\times10^{6}$ GHz) imaging polarimetric observations at an angular resolution of $0\farcs1$ ($6$ pc) with the Faint Object Camera (FOC) aboard the \textit{Hubble Space Telescope} (\textit{HST}). The \textit{HST}/FOC observations have been reported by \citet{Kishimoto1999}. 

Here, we present the re-reduced observations by \citet{Barnouin}, which provides data with higher signal-to-noise than those previously produced by \citet{Kishimoto1999}. This is the first dataset of a large program to re-calibrate the imaging polarimetric observations of active nuclei observed by \textit{HST}/FOC. Figure \ref{fig:fig1} presents the total intensity maps of the sub-mm, MIR, and UV observations used in this work.

We perform aperture photometry of the radio knots as follows. We use a circular aperture equal to the lowest angular resolution of the sub-mm-UV observations, which is $0\farcs3$ ($18$ pc) from the CanariCam/GTC observations. We sum the fluxes within the aperture and subtract the  background level within the aperture. The background level is estimated using a region of the field-of-view from the observations without emission from the source. 
The final photometric error is estimated as the sum in quadrature of the background and photometric calibrations associated to each observations. Table \ref{tab:table1} shows the measured photometry per wavelength.

Although the MIR observations show an extended point source (Fig. \ref{fig:fig1}, middle panel), our measured fluxes are dominated, $>90$\%, by the central unresolved source, where $<10$\% of the total flux may arise from diffuse emission in the host galaxy \citep{2006ApJ...640..612M, LR2018}. Furthermore, the resolved total emission at $2~\mu$m using VLTI/GRAVITY and $8-12~\mu$m using VLTI/MATISSE have been found to arise from optically thin dust in the central $<0.5$ pc above the inner edge of the dusty torus \citep{2022Natur.602..403G}. The midplane of the dusty torus is optically thick within the $2-10~\mu$m wavelength range \citep{LR2015,LR2018,2022Natur.602..403G}. These results constrain the MIR emission to arise from the central $0.1-0.5$ pc of the dusty torus of NGC~1068.

We estimate the B-field strength of the knots as follows. For the `S1' knot, the B-field strength in the resolved dusty torus has been previously estimated using the $860~\mu$m imaging polarimetric observations with ALMA \citep{2020ApJ...893...33L}. The ALMA observations measured the polarized flux arising from thermal emission by means of magnetically aligned dust grains in the equatorial plane of the dusty torus. Using the Davis-Chandrasekhar-Fermi (DCF) method \citep{Davis1951,CF1953}, the B-field strength was estimated to be $0.67^{+0.94}_{-0.31}$ mG in the $3–8$ pc region of the eastern side along the equatorial plane of the torus. Furthermore, near-infrared (NIR; $2.2~\mu$m) imaging polarimetric observations using MMT/MMT-pol  measured the polarization from the central core of NGC~1068 at an angular resolution of $0\farcs2$ ($12$ pc) \citep{LR2015}. The strong intrinsic polarization level of $7.0\pm2.2$\% with a position angle of polarization of $127^{\circ}$ indicated the presence of a strong and ordered B-field. The polarization arises from magnetically aligned dust grains from hot dust, $T_{\rm{d}} \sim 800-1500$ K, at the inner edge of the dusty torus. Using a modified version of the DCF method, to account for no equipartition between the kinetic and magnetic energy, the B-field strength was estimated to be $139^{+11}_{-20}$ mG at $0.4$ pc of the dusty torus. Assuming that the B-field strength decreases with the distance from the core as $B \propto r^{-1}$, the B-field strength is estimated to be $B\sim700-300$ mG at the sublimation radius, $r_{\rm{sub}} = 0.1-0.2$ pc, of the dusty torus. We take a fiducial B-field strength of $B_{\rm{S1}}=500$ mG for a radius of $R_{\rm{S1}}=0.1$ pc. 

The knots `C' and `NE' are dominated by synchrotron emission at sub-mm wavelengths \citep{2019A&A...632A..61G}. The $860~\mu$m ALMA polarimetric observartions detected polarization levels of up to $\sim7$\% and $\sim11$\% in the `C' and `NE' knots respectively. For both knots, the B-field orientations seems to be related to the  shock front between the jet and the GMC in the ISM \citep{2020ApJ...893...33L} and heavily depolarized due to Faraday depolarization. We estimate the minimum magnetic field strength as $B_{\rm{min}} = 1.8\times10^{4} (\eta L_{\nu}) / V)^{2/7} \nu^{1/7}$ G, where $\eta$ is the fraction of electrons contributing to the total energy, $L_{\nu}$ is the luminosity  in Watt at the observed frequency $\nu$ in Hz, and $V$ is the volume of the source in m$^{3}$. This equation assumes equipartition between B-field and relativistic particles and a spectral index of $\alpha=0.75$. The spectral indices of the `S1' and `'NE' knots are found to be $\sim0.79$ \citep{2019A&A...632A..61G}. Taking $\eta=2\times10^{3}$ (i.e., electrons emit all the energy), the $860~\mu$m fluxes in Table \ref{tab:table1}, physical sizes of $R_{\rm{C}}=0.1$ pc, and a distance of $14.4$ Mpc, we estimate the minimum B-field strength of knot C and NE as $B_{\rm{min,C}} = 83$ mG and $B_{\rm{min,NE}} = 68$ mG, respectively. Note that the region emitting synchrotron emission is highly localized and well-below the angular size of the radio observations \citep{2004ApJ...613..794G}, thus we assume an upper-limit of R$_{\rm{C}}=0.1$ pc.

The SEDs of knots `S1' and `C' are shown in Figure~\ref{fig:sed}. Due to the relatively low $\gamma$-ray opacity at knot `NE', it is unlikely a strong high-energy neutrino emitter and thus not included in the SED plot. However, we present the physical properties of knot `NE' in this section for completeness.

\section{Neutrino and gamma-ray production}\label{sec:nu}
Below we focus on knot S1 and knot C as potential neutrino emission sites. As explained in the previous section, we adopt benchmark parameters $B=500$~mG and $R = 0.1$~pc for both knots and discuss the effect of alternative parameter values in Section~\ref{sec:dis}. The radiation field at knot NE is too weak to attenuate the sub-TeV $\gamma$-rays co-produced with high-energy neutrinos and therefore cannot be an effective neutrino production site.

\begin{figure}[t!]
    \centering
   \includegraphics[width=0.49\textwidth]{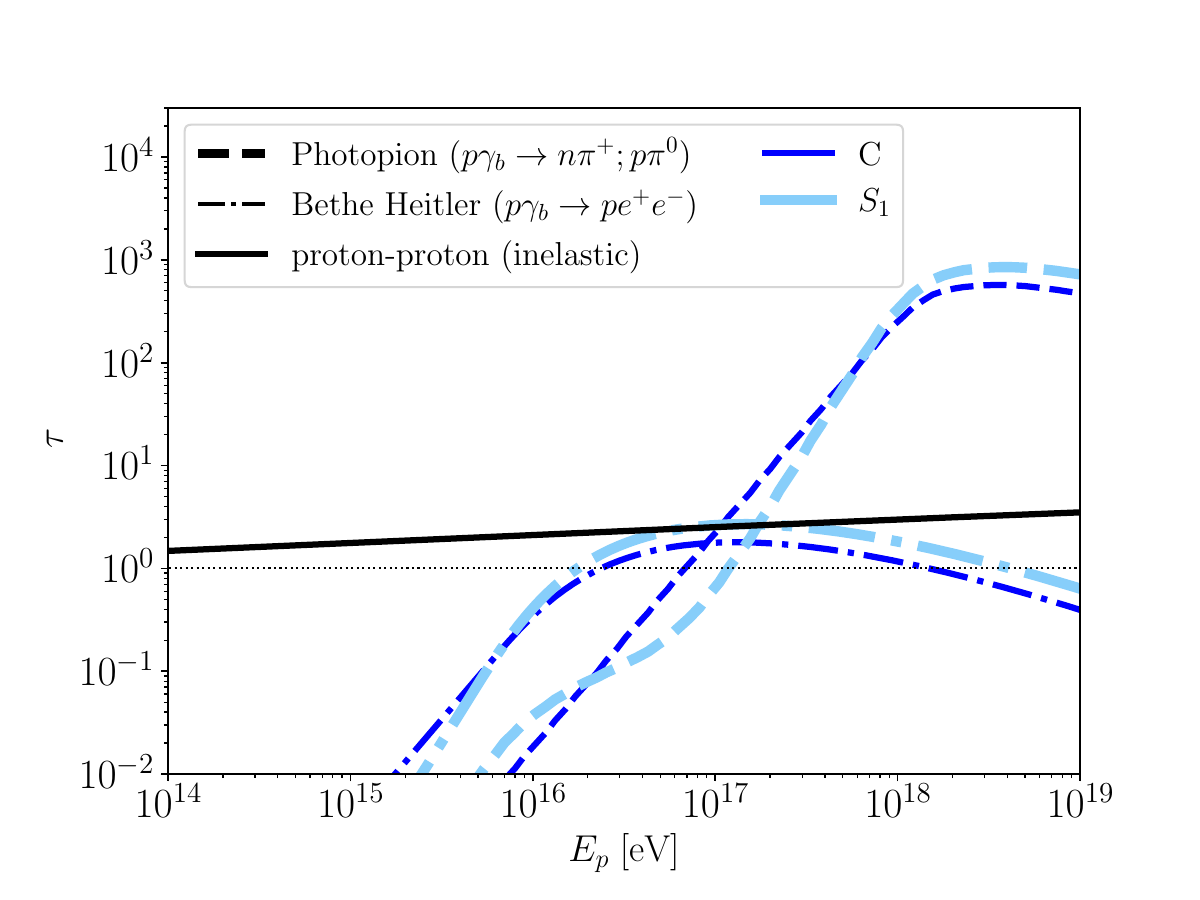}
    \caption{
    \label{fig:tau_p} Inelastic optical depth to protons due to proton-proton interaction (black solid curve), photopion production (colored dashed curves), and Bethe-Heitler process (colored dash-dotted curves) as a function of the proton energy assuming $R=0.1$~pc, $n_{\rm gas}=10^5\,\rm cm^{-3}$ and $\beta_{\rm sh} = 10^{-3}$ for S1 (light blue) and C (dark blue) knots, respectively. For reference, the thin dotted line shows $\tau = 1$.} 
\end{figure}

\subsection{Neutrino production}
Particle acceleration may happen in the accretion outflows and shocks produced by the jet-ISM collisions. Assuming that the radio knots have a physical size of $R=0.1\,R_{-1}$~pc and magnetic field $B=0.5\,B_{-0.3}$~G, protons may be accelerated up to $E_{p,\rm max}=4.7\times 10^{16}\,\beta_{-3}\eta_{\rm acc,-1}\,B_{-0.3}\,R_{-1}\,\rm eV$, where $\eta_{\rm acc}\sim 0.1$ is the fraction of the shock energy density that is injected into cosmic rays, i.e., the acceleration efficiency,  and $\beta=v_{\rm sh}/c= 10^{-3}\,\beta_{-3}$ \citep{Axon_1998, Roy:2000dm} is the velocity of the diffusive shocks or outflows. 
 
ALMA observations suggest an $H_2$ density of $n_{\rm gas} \sim 10^5-10^7\,\rm cm^{-3}$ in the torus and knots \citep{2019A&A...632A..61G}. 
The effective optical depth of the molecular gas to proton-proton (pp) interaction is 
\begin{equation}
    \tau_{\rm pp} \approx \frac{t_{\rm conf}}{t_{\rm pp}} \approx 1.5 \,R_{-1} \beta_{\rm sh, -3}^{-1} n_{\rm gas, 5},
\end{equation}
where $t_{\rm conf} = \min(t_{\rm diff}, t_{\rm dyn})$ is the time when the protons are confined, $t_{\rm dyn} = R/ v_{\rm sh}$ is the dynamical time, $t_{\rm conf}\sim R^2/D$ is the diffusion time of cosmic rays, $D\approx (B/\delta B)^2\, c r_g /3 $ is the diffusion coefficient in the Bohm limit \citep{2006ApJ...642..902P}, $r_g=E_p / eB$ is the Larmor radius, $\delta B$ is the amplitude of random field. The pp interaction time is $t_{\rm pp}\sim (n_{\rm gas}\sigma_{\rm pp}\kappa_{\rm pp} c)^{-1}$, with $\sigma_{\rm pp}\kappa_{\rm pp}= 4.8\times 10^{-26}\,\rm cm^2$ being the inelastic cross section of the interaction of a proton at $100$~TeV with a rest-mass proton \citep{2014PhRvD..90l3014K}.

In the jet-ISM interaction region, the photopion production ($p\gamma_b\rightarrow n \pi^+$ and $p\gamma_b\rightarrow p\pi^0$) and Bethe-Heitler process ($p\gamma\rightarrow p e^+ e^-$) are ineffective for TeV-PeV protons. The energy spectra of the radiation fields at the knots peak at IR energies, which are too low to interact with the protons that produce the IceCube neutrinos. As shown in Figure~\ref{fig:tau_p}, these processes may be relevant for cosmic rays above $\sim$50~PeV, if particles at such high energies may be accelerated in the outflows.    

We assume that the proton spectrum follows a simple power law, $dN/dE_p \propto E_p^ {-s}$, with $s = 1$ up to a break energy of 500~TeV and $s = 3.2$ above the break. The shape of the proton spectrum above the break energy barely impacts the neutrino and $\gamma$-ray spectra below $\sim 10$~TeV. 
A hard proton spectrum may be caused by several factors. First, when the shock acceleration efficiency is high, an increase in the shock compression ratio due to the presence of relativistic particles may yield a spectral index smaller than 2 \citep{1984A&A...132...97A}. Second, such a hard spectrum may be caused by the escape of the highest-energy particles ahead of the shock front \citep{2005MNRAS.361..907B}. Finally, when accelerated cosmic rays penetrate a dense gas clump, the higher-energy particles penetrate more efficiently as a result of a larger diffusion coefficient. This would also lead to a harder spectrum than at the shock \citep{2019MNRAS.487.3199C}. In addition to diffusive shock acceleration, other mechanisms may also produce a hard proton spectrum. A proton spectrum with $s\gtrsim 1$ would produce too much GeV-TeV $\gamma$-ray emission and be inconsistent with the $\gamma$-ray observations.

\subsection{Optical depth to $\gamma$-rays}

\begin{figure} [t]
    \centering
   \includegraphics[width=0.49\textwidth]{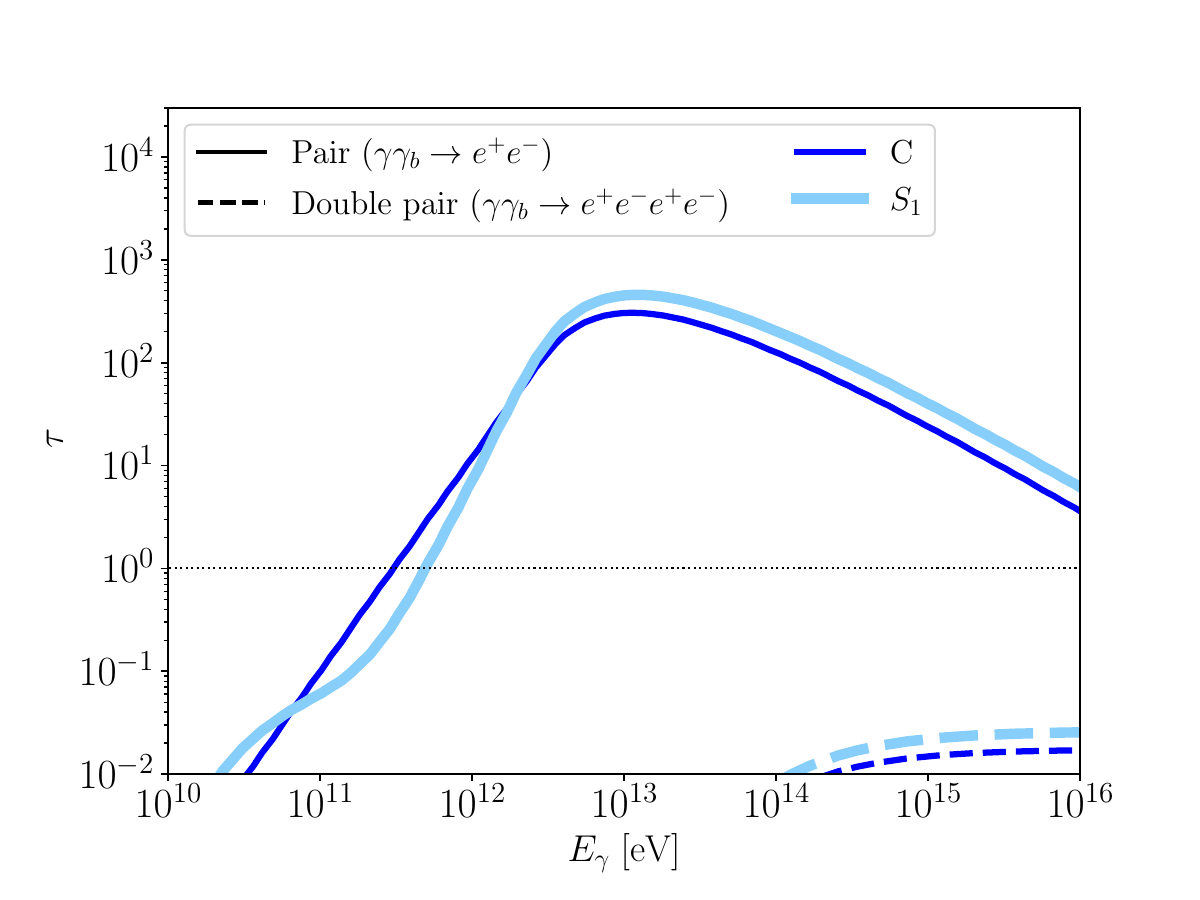}
    \caption{
    \label{fig:tau_gamma} Optical depth to $\gamma$-rays due to pair (solid curves) and double pair production (dashed curves) as a function of the $\gamma$-ray energy assuming $R=0.1$~pc for S1 (light blue) and C (dark blue) knots, respectively. As in the previous plot, the thin dotted line corresponds to $\tau = 1$.    } 
\end{figure}

\begin{figure*} 
    \centering
   \includegraphics[width=0.9\textwidth]{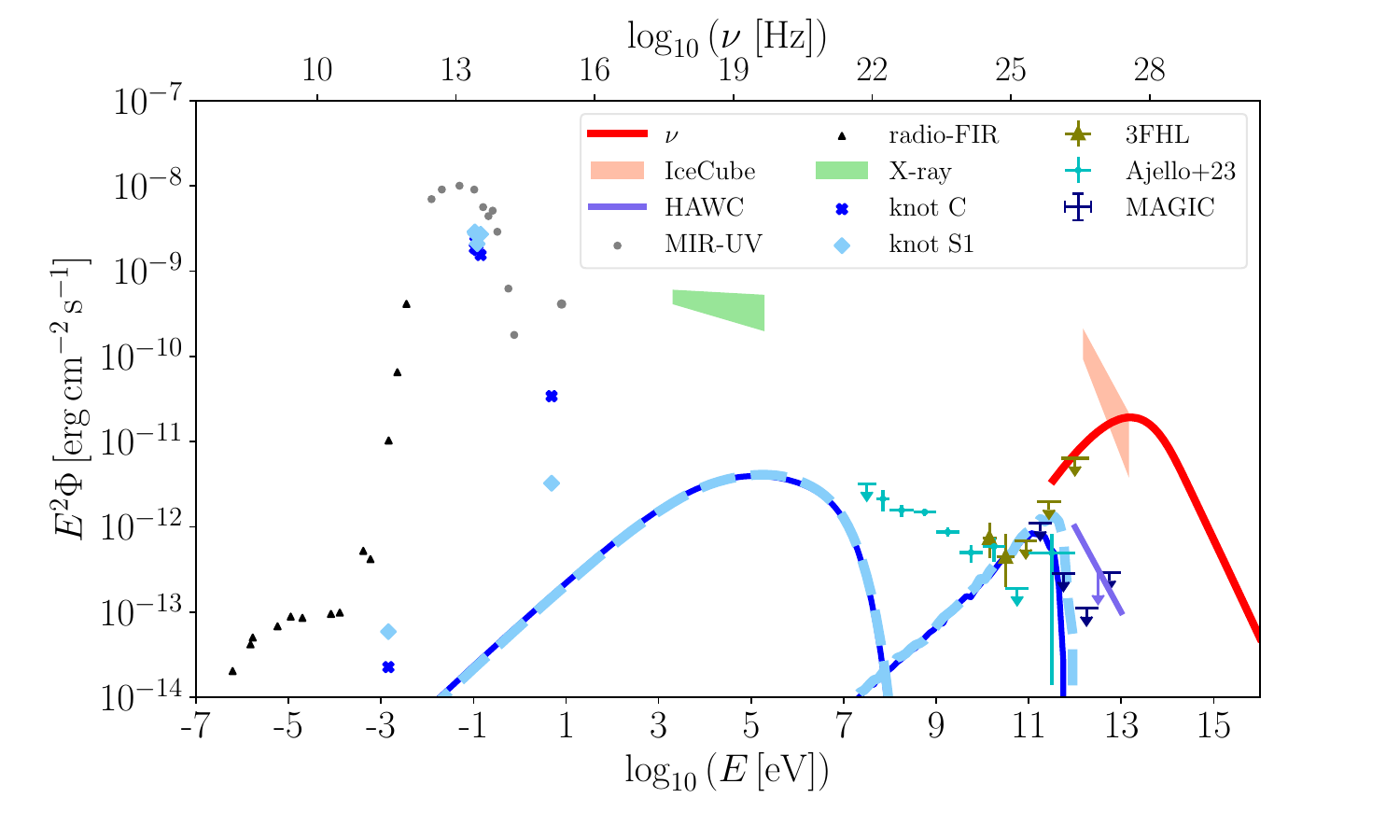}
    \caption{
    \label{fig:sed} Broadband spectral energy distribution of NGC~1068 and its knots C (in dark blue) and S1 (in light blue) in the inner circumnuclear region. The red curve indicates the all-flavor high-energy neutrino emission produced by one knot (assumed to be the same for both knots). The curves (solid for knot C and dashed for knot S1) show the synchrotron and inverse Compton radiation by the accompanying $\gamma$-rays and their electromagnetic cascades, respectively. 
    The model parameters in use are  $R=0.1$~pc and $B=500$~mG.  The radio to UV measurements of the two knots, from Section~\ref{sec:knotObs}, are indicated by cross and diamond markers, respectively. The rest of the data points show the multi-messenger emission of NGC~1068. In particular, $\gamma$-ray data points and upper limits are from observations of {\it Fermi}-LAT, including 3FHL \citep{2017ApJS..232...18A} catalog and \citet{2023arXiv230702333A}, MAGIC \citep{2019ApJ...883..135A}, and HAWC (assuming $E^{-3}$ spectrum, \citealt{2022ATel15765....1W}). The band in X-ray corresponds to the  nucleus component of the best-fit model ``M2d" of \citet{2015ApJ...812..116B}, which shows the intrinsic, unabsorbed flux of the nucleus at 2-195~keV.
    The radio, IR, and optical data points are based on observations of NGC~1068 by \citet{2009ApJ...704.1433M, 2013MNRAS.434..956C, 2016A&A...594A..26P, 2017A&A...598A..78I}, \citet{2006ApJ...640..612M}, and \citet{2005MNRAS.361...34D, 2006AJ....131.1163S, 2007ApJS..173..185G} respectively. 
}
\end{figure*}

Based on the spectral energy distribution of the radiation fields found in Section~\ref{sec:knotObs}, we compute the optical depth to high-energy $\gamma$-rays due to pair production ($\gamma\gamma_b\rightarrow e^+e^-$) and double pair production ($\gamma\gamma_b\rightarrow e^+e^-e^+e^-$) processes. As shown in Figure~\ref{fig:tau_gamma}, with a size $R=0.1$~pc, knots S1 and C are optically thick to $\gamma$-rays above $\sim$400~GeV and $\sim$200~GeV, respectively. The TeV $\gamma$-rays produced together with the neutrinos, therefore, would be attenuated by the low-energy photons. 

The magnetic energy density at the knots dominates over the radiation energy density. 
The intensity of knot~C at $\varepsilon=0.1$~eV corresponds to a differential luminosity of $\varepsilon L_\varepsilon = 4.6\times 10^{43}\,\rm erg\,s^{-1}$ for a source distance of 14.4~Mpc \footnote{A high internal photon energy density in an optically-thick source could be due to a high dust temperature obscured by the foreground \citep{2006A&A...446..813G, 2019MNRAS.484.3665Y}.}.  The ratio of the magnetic energy density and radiation energy density is $w_B / w_{\gamma_b} = 7.8\,B_{-0.3}^2$. Electrons from the pair production would dissipate most of their energy through synchrotron radiation. A small fraction of them would up-scatter the IR and optical photons to 10-100~GeV $\gamma$-rays.

\subsection{Gamma-ray and neutrino spectra}

We compute the neutrino and injected $\gamma$-ray spectra using proton-proton cross sections from \citet{2021PhRvD.104l3027K} above 4~GeV and \citet{2006PhRvD..74c4018K} below 4~GeV. The electromagnetic cascades of $\gamma$-rays are computed using a Monte Carlo code based on the thinning technique of {\it CRPropa~3.2} \citep{2022JCAP...09..035A}. 

Figure~\ref{fig:sed} presents the broad-band spectral energy distribution of NGC~1068. Our model may explain $\sim 20\%$ of the neutrino flux at 3~TeV and $\sim 100\%$ of the flux above 10~TeV. The total neutrino luminosity is $L_\nu = 1.9\times 10^{42}\,\rm erg\,s^{-1}$ above 1~TeV. Since the jet-ISM interaction region is optically thin to $\gamma$-rays below $\sim 200$~GeV, the neutrino emission by the knots is constrained by the non-detection of the accompanying $\gamma$-ray emission at 0.1-1~TeV.

The $\gamma$-rays at $\sim 3-200$~GeV are mostly unattenuated pion decay products. Therefore, they follow the injection spectrum of $dN/dE_\gamma$, which has a similar shape as the proton spectrum $dN/dE_p\propto E_p^{-1}$. Electrons from the pair production of $\gamma$-rays are quickly cooled due to synchrotron radiation and inverse Compton scattering. The cooling results an electron spectrum that follows roughly $dN/dE_e\propto E_e^{-2}$ and a corresponding synchrotron spectrum of $dN/dE_\gamma\propto E_\gamma^{-1.5}$ up to a peak energy around $\sim0.1$~MeV. Since the injected proton spectra in the models of both knots are assumed to be the same and the synchrotron emission is dominated by the injected power, the synchrotron spectra of the two knots look identical. 

Our model predicts a peak in the $\gamma$-ray energy spectrum at 100~GeV to 1~TeV. Such a peak cannot be produced by proton interactions in the corona or starburst activities at larger radius and therefore is a unique feature of the jet-ISM interaction region.   The peak is consistent with the analyses of the latest {\it Fermi}-LAT data, which suggest that NGC~1068 is marginally detected at 0.1-1~TeV at test statistic ${\rm TS} \sim 8$ \citep{2023arXiv230702333A, 2023arXiv230703259B}.
Deeper observations of   imaging air Cherenkov telescopes (IACTs) in the future may reveal or further constrain such a component.

\section{conclusions and discussion}\label{sec:dis}
One of the first high-energy neutrino sources, NGC~1068, turned out to be highly obscured to TeV $\gamma$-rays. We show that protons may be accelerated by shocks generated when the jet collides with molecular clouds in the circumnuclear region, interact with the gas, and produce high-energy neutrinos and gamma-rays. Based on multi-wavelength observations of the knots and numerical simulation of electromagnetic cascades, we find that $\gamma$-rays above $\sim$200~GeV energies are attenuated due to interaction with IR and optical radiation fields. Comparing to the source spectrum measured by \citet{IceCubeNGC1068} using a single power-law, the jet-ISM interaction region may explain the $\sim100$\% of the observed neutrino flux above $\sim$10~TeV and contribute to 20\% of the observed flux at 3~TeV. 

Jet-ISM interaction is commonly observed in both Galactic \citep{HAWC:2018gwz} and extragalactic jets. Notably, collision of jetted material is also evident in other candidate neutrino sources, such as TXS~0506+056 \citep{2019A&A...630A.103B} and NGC~4151 \citep{2011ApJ...736...62W}.  Our model suggest that such regions may be promising sites for high-energy neutrino production. The thermal radiation at the interaction site may attenuate the $\gamma$-rays accompanying the neutrinos. $\gamma$-ray-obscured sources are needed to explain the diffuse astrophysical neutrino flux due to the tension of the cascaded $\gamma$-ray flux and the isotropic $\gamma$-ray background measured by {\it Fermi}-LAT \citep{2016PhRvL.116g1101M, 2022ApJ...933..190F}. Jet-ISM interaction regions like NGC~1068's knots may serve as such $\gamma$-ray-hidden sources.

\citet{2022arXiv220702097I} studied the neutrino production in the outer torus region and found a relatively low neutrino flux. The model presented in this work is different in two aspects. First, our model focuses on the regions where the jet interacts with the circumnuclear region rather than the torus itself. The radiation field revealed by our observation extends to optical and UV bands, which help attenuate $\gamma$-rays at TeV energies. Second, the magnetic field in the knots is observed to be significantly higher than that in the outer torus region assumed by \citet{2022arXiv220702097I}. 
Thus most of the pairs dissipate their energy through synchrotron radiation at 0.1-1~MeV instead of inverse Compton radiation at 0.1-1~TeV. 

The proton luminosity in our model is $L_p\sim 2\, L_\nu =3.8\times 10^{42}\,\rm erg\,s^{-1}$. We have assumed a hard proton spectrum with index $s = 1$. If the particles were accelerated with a softer spectrum, such as $s \sim 2-3$, the power of the relativistic protons would have to be significantly lower to be consistent with the $\gamma$-ray constraints at 1-10~TeV. In that case, the neutrino flux would be negligible. The jet-ISM interaction would instead contribute to the $\gamma$-ray emission at 1-100~GeV, which can hardly be explained by star formation regions alone \citep{2014ApJ...780..137Y}.

Our benchmark model adopts $B=500$~mG.  While an even higher field is possible at knot S1, in general $B\gtrsim 300$~mG allows $w_B\gtrsim w_{\gamma_b}$ and thus yields similar results. A magnetic field of $B\lesssim 300$~mG or an emission region with size $R\gtrsim 0.2$~pc would cause $w_B \ll w_{\gamma_b}$ and hence overproduce sub-TeV $\gamma$-rays. 
As the $\gamma$-ray attenuation sensitively depends on the strength of the magnetic field and the size of the emission region, high angular resolution IR-sub-mm polarimetric observations are crucial to resolving a $\gamma$-ray-opaque neutrino emission site.

Deeper observations of NGC~1068 by IACTs and future data from {\it Fermi}-LAT at $0.1-1$~TeV may reveal the sub-TeV $\gamma$-ray flux predicted by our model or further constrain the opacity of the neutrino emission site to high-energy $\gamma$-rays. Future observation by IceCube and next-generation neutrino telescopes may also better measure the neutrino spectral shape and resolve the contribution by various components of NGC~1068.

\begin{acknowledgments}
We thank Roger Blandford for helpful comments on the manuscript. 
The work of K.F and F.H is supported by the Office of the Vice Chancellor for Research and Graduate Education at the University of Wisconsin-Madison with funding from the Wisconsin Alumni Research Foundation. K.F. acknowledges support from National Science Foundation (PHY-2110821, PHY-2238916) and NASA (NMH211ZDA001N-Fermi). J.S.G. thanks the University of Wisconsin College of Letters and Science for partial support of his IceCube-related research. The research of F.H was also supported in part by the U.S. National Science Foundation under grants~PHY-2209445 and OPP-2042807.

This paper makes use of the following ALMA data: ADS/JAO.ALMA\#2016.1.00176.S. ALMA  is a partnership of ESO (representing its member states), NSF (USA) and NINS (Japan), together with NRC (Canada), MOST and ASIAA (Taiwan), and KASI (Republic of Korea), in  cooperation with the Republic of Chile. The Joint ALMA Observatory is operated by  ESO, AUI/NRAO and NAOJ. The National Radio Astronomy Observatory is a facility of the National Science Foundation operated under cooperative agreement by Associated Universities, Inc.
\end{acknowledgments}

\bibliography{ref}

\end{document}